\documentclass[a4paper]{article}
\usepackage{INTERSPEECH2019}
\usepackage{amsmath,graphicx,amssymb}
\usepackage[export]{adjustbox}
\usepackage{subfig}
\usepackage{color}
\usepackage{enumitem}
\usepackage{multirow}
\PassOptionsToPackage{hyphens}{url}\usepackage{hyperref}

\hypersetup{colorlinks=true}

\title{VoiceFilter: Targeted Voice Separation by \\ Speaker-Conditioned Spectrogram Masking}

\name{\begin{tabular}{c}
Quan Wang\textsuperscript{* 1} \quad Hannah Muckenhirn\textsuperscript{* 2,3} \quad Kevin Wilson\textsuperscript{1} \quad Prashant Sridhar\textsuperscript{1} \\
Zelin Wu\textsuperscript{1} \quad John Hershey\textsuperscript{1} \quad Rif A. Saurous\textsuperscript{1} \quad Ron J. Weiss\textsuperscript{1} \quad Ye Jia\textsuperscript{1} \quad Ignacio Lopez Moreno\textsuperscript{1}
\thanks{\hspace{-1mm}* Equal contribution. Hannah performed this work as an intern at Google.}
\end{tabular}}

\address{
\textsuperscript{1}Google Inc., USA \qquad
\textsuperscript{2}Idiap Research Institute, Switzerland \qquad
\textsuperscript{3}EPFL, Switzerland}

\email{
  \normalsize
  \href{mailto:quanw@google.com}{\nolinkurl{quanw@google.com}}
  \qquad
  \href{mailto:hannah.muckenhirn@idiap.ch}{\nolinkurl{hannah.muckenhirn@idiap.ch}}
}

\begin{document}

\maketitle

\begin{abstract}
In this paper, we present a novel system that separates the voice of a target speaker from multi-speaker signals, by making use of a reference signal from the target speaker. We achieve this by training two separate neural networks: (1) A speaker recognition network that produces speaker-discriminative embeddings; (2) A spectrogram masking network that takes both noisy spectrogram and speaker embedding as input, and produces a mask. Our system significantly reduces the speech recognition WER on multi-speaker signals, with minimal WER degradation on single-speaker signals.
\end{abstract}


\vspace{-5pt}
\section{Introduction}
\label{sec:intro}
Recent advances in speech recognition have led to performance improvement in challenging scenarios such as noisy and far-field conditions. However, speech recognition systems still perform poorly when the speaker of interest is recorded in crowded environments, \textit{i.e.}, with interfering speakers in the foreground or background.

One way to deal with this issue is to first apply a \emph{speech separation} system on the noisy audio in order to separate the voices from different speakers. Therefore, if the noisy signal contains $N$ speakers, this approach would yield $N$ outputs with a potential additional output for the ambient noise. A classical speech separation task like this needs to cope with two main challenges. First, identifying the number of speakers $N$ in the recording, which in realistic scenarios is unknown. Secondly, the optimization of a speech separation system may be required to be invariant to the permutation of speaker labels, as the order of the speakers should not have an impact during training~\cite{hershey2016deep}. Leveraging the advances in deep neural networks, several successful works have been introduced to address these problems, such as deep clustering~\cite{hershey2016deep}, deep attractor network~\cite{chen2017deep}, and permutation invariant training~\cite{yu2017permutation}. 

This work addresses the task of isolating the voices of a subset of speakers of interest from the commonality of all the other speakers and noises. For example, such subset can be formed by a single target speaker issuing a spoken query to a personal mobile device, or the members of a house talking to a shared home device. We will also assume that the speaker(s) of interest can be individually characterized by previous reference recordings, \emph{e.g.} through an enrollment stage. This task is closely related to classical speech separation, but in a way that it is speaker-dependent. In this paper, we will refer to the task of speaker-dependent speech separation as \emph{voice filtering} (some literature call it \emph{speaker extraction}). We argue that for voice filtering, speaker-independent techniques such as those presented in~\cite{hershey2016deep, chen2017deep, yu2017permutation} may not be a good fit. In addition to the challenges described previously, these techniques require an extra step to determine which output -- out of the $N$ possible outputs of the speech separation system -- corresponds to the target speaker(s), by \textit{e.g.} choosing the loudest speaker, running a speaker verification system on the outputs, or matching a specific keyword. 

A more end-to-end approach for the voice filtering task is to treat the problem as a binary classification problem, where the positive class is the speech of the speaker of interest, and the negative class is formed by the combination of all foreground and background interfering speakers and noises. By speaker-conditioning the system, this approach suppresses the three aforementioned challenges: unknown number of speakers, permutation problem, and selection from multiple outputs. In this work, we aim to condition the system on the speaker embedding vector of a reference recording. The proposed approach is the following. We first train a LSTM-based speaker encoder to compute robust speaker embedding vectors. We then train separately a time-frequency mask-based system that takes two inputs: (1) the embedding vector of the target speaker, previously computed with the speaker encoder; and (2) the noisy multi-speaker audio. This system is trained to remove the interfering speakers and output only the voice of the target speaker.\footnote{Samples of output audios are available at: \url{https://google.github.io/speaker-id/publications/VoiceFilter}}
This approach can be easily extended to more than one speaker of interest by repeating the process in turns, for the reference recording of each target speaker.

Similar related literature exists for the task of voice filtering. For example, in~\cite{ephrat2018looking,afouras2018conversation}, the authors achieved impressive results by doing an indirect speaker conditioning of the system on the visual information (lips movement). However, a solution like that would require simultaneously using speech and visual information, which may not be available in certain type of applications, where a reference speech signal may be more practical.
The systems proposed in ~\cite{zmolikova2017speaker,wang2018deep,vzmolikova2017learning,delcroix2018single} are also very similar to ours, with a few major differences: (1) Unlike using one-hot vectors, i-vectors or speaker posteriors derived from a cross-entropy classification network, our speaker encoder network is specifically designed for large-scale end-to-end speaker recognition~\cite{ge2e}, which proves to perform much better in speaker recognition tasks, especially for unseen speakers. (2) Instead of using a GEV beamformer~\cite{zmolikova2017speaker,vzmolikova2017learning}, our system directly optimizes the power-law compressed reconstruction error between the clean and enhanced signals~\cite{wilson2018exploring}. (3) In addition to source-to-distortion ratio~\cite{zmolikova2017speaker,wang2018deep}, we focus on Word Error Rate improvements for ASR systems. (4) We use dilated convolutional layers to capture low-level acoustic features more effectively. (5) We prefer separately trained speaker encoder network over joint training like~\cite{vzmolikova2017learning,delcroix2018single}, due to the very different requirements for data in speaker recognition and source separation tasks.


The rest of this paper is organized as follows. In Section~\ref{sec:models}, we describe our approach to the problem, and provide the details of how we train the neural networks. In Section~\ref{sec:exp}, we describe our experimental setup, including the datasets we use and the evaluation metrics. The experimental results are presented in Section~\ref{sec:results}. We draw our conclusions in Section~\ref{sec:conclusions}, with discussions on future work directions.

\begin{figure*}
  \centering
    \includegraphics[width=0.87\textwidth]{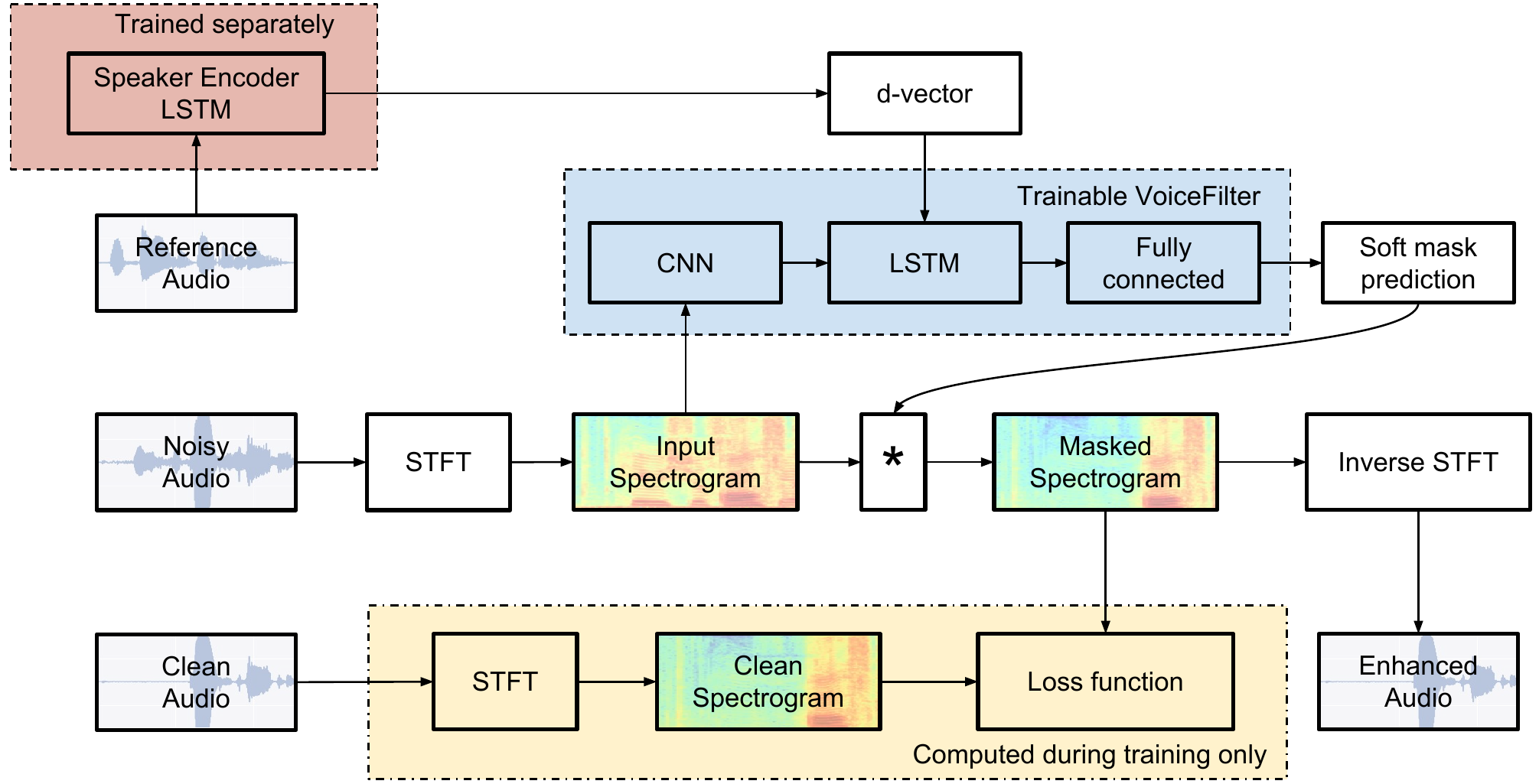}
  \caption{System architecture.}
  \label{fig:system}
  \vspace{-10pt}
\end{figure*}

\vspace{-5pt}
\section{Approach}
\label{sec:models}

The system architecture is shown in Fig. \ref{fig:system}. The system consists of two separately trained components: the speaker encoder (in red), and the VoiceFilter system (in blue), which uses the output of the speaker encoder as an additional input. In this section, we will describe these two components.   

\subsection{Speaker encoder}

The purpose of the speaker encoder is to produce a speaker embedding from an audio sample of the target speaker. This system is based on a recent work from Wan \textit{et al.} \cite{ge2e}, which achieves great performance on both text-dependent and text-independent speaker verification tasks, as well as on speaker diarization \cite{wang2017speaker,zhang2018supervised},  multispeaker TTS \cite{jia2018transfer}, and speech-to-speech translation \cite{jia2019direct}.

The speaker encoder is a 3-layer LSTM network trained with the generalized end-to-end loss \cite{ge2e}. It takes as inputs log-mel filterbank energies extracted from windows of 1600 ms, and outputs speaker embeddings, called \emph{d-vectors}, which have a fixed dimension of 256. To compute a d-vector on one utterance, we extract sliding windows with 50\% overlap, and average the L2-normalized d-vectors obtained on each window.

\subsection{VoiceFilter system}
The VoiceFilter system is based on the recent work of Wilson \textit{et al.} \cite{wilson2018exploring}, developed for speech enhancement. As shown in Fig.~\ref{fig:system}, the neural network  takes two inputs: a d-vector of the target speaker, and a \emph{magnitude spectrogram} computed from a noisy audio. The network predicts a soft mask, which is element-wise multiplied with the input (noisy) magnitude spectrogram to produce an enhanced magnitude spectrogram. To obtain the enhanced waveform, we directly merge the phase of the noisy audio to the enhanced magnitude spectrogram, and apply an inverse STFT on the result.
The network is trained to minimize the difference between the masked magnitude spectrogram and the target magnitude spectrogram computed from the clean audio.

The VoiceFilter network is composed of 8 convolutional layers, 1 LSTM layer, and 2 fully connected layers, each with ReLU activations except the last layer, which has a sigmoid activation. The values of the parameters are provided in Table~\ref{tab:architecture}.
The d-vector is \mbox{\textbf{repeatedly concatenated}} to the output of the last convolutional layer in every time frame. The resulting concatenated vector is then fed as the input to the following LSTM layers. We decide to inject the d-vector between the convolutional layers and the LSTM layer and not before the convolutional layers for two reasons. First, the d-vector is already a compact and
robust representation of the target speaker, thus we do not need to modify it by applying convolutional layers on top of it. Secondly, convolutional layers assume time and frequency homogeneity, and thus cannot be applied on an input composed of two completely different signals: a magnitude spectrogram and a speaker embedding.

\begin{table}
\centering
  \caption{Parameters of the VoiceFilter network.}
  \label{tab:architecture}
  \vspace{-2mm}
  \begin{tabular}{| c | c | c|  c | c |c|}
    \hline
    \multirow{2}{*}{\bf Layer} & \multicolumn{2} {c|} {\bf Width} & \multicolumn{2} {c|} {\bf Dilation} & \multirow{2}{*}{\bf Filters / Nodes} \\ \cline{2-5} 
    & time & freq & time & freq &\\ \hline
    CNN 1 &1&7&1&1&64\\
    CNN 2 &7&1&1&1&64\\
    CNN 3 &5&5&1&1&64\\
    CNN 4 &5&5&2&1&64\\
    CNN 5 &5&5&4&1&64\\
    CNN 6 &5&5&8&1&64\\
    CNN 7 &5&5&16&1&64\\
    CNN 8 &1&1&1&1&8\\
    LSTM &-&-&-&-&400\\
    FC 1 &-&-&-&-&600\\
    FC 2 &-&-&-&-&600\\ \hline
  \end{tabular}
  \vspace{-15pt}
\end{table}

While training the VoiceFilter system, the input audios are divided into segments of 3 seconds each and are converted, if necessary, to single channel audios with a sampling rate of 16 kHz.

\vspace{-5pt}
\section{Experimental setup}
\label{sec:exp}
In this section, we describe our experimental setup: the data used to train  the two components of the system separately, as well as the metrics to assess the systems.

\subsection{Data}
\subsubsection{Datasets}
\paragraph*{Speaker encoder:} Although our speaker encoder network has exactly the same network topology as the text-independent model described in \cite{ge2e}, we use more training data in this system. Our speaker encoder is trained with two datasets combined by the MultiReader technique introduced in \cite{ge2e}. The first dataset consists of anonymized voice query logs in English from mobile and farfield devices. It has about 34 million utterances from about 138 thousand speakers. The second dataset consists of LibriSpeech \cite{panayotov2015librispeech}, VoxCeleb \cite{nagrani2017voxceleb}, and VoxCeleb2 \cite{chung2018voxceleb2}. This model (referred to as ``d-vector V2'' in \cite{zhang2018supervised}) has a 3.06\% equal error rate~(EER) on our internal en-US phone audio test dataset, compared to the 3.55\% EER of the one reported in \cite{ge2e}.

\vspace{-10pt}

\paragraph*{VoiceFilter:} We cannot use a ``standard'' benchmark corpus for speech separation,
such as one of the CHiME challenges~\cite{barker2017chime}, because  we need a clean reference utterance of each target speaker in order to compute speaker embeddings. Instead, we train and evaluate the VoiceFilter system on our own generated data, derived either from the VCTK dataset~\cite{veaux2016superseded} or from the LibriSpeech dataset~\cite{panayotov2015librispeech}. For VCTK, we randomly take 99 speakers for training, and 10 speakers for testing. For LibriSpeech, we used the training and development sets defined in the protocol of the dataset: the training set contains 2338 speakers, and the development set contains 73 speakers. These two datasets contain read speech, and each utterance contains the voice of one speaker. We explain in the next section how we generate the data used to train the VoiceFilter system.

\subsubsection{Data generation}
\begin{figure}
  \centering
    \includegraphics[width=0.9\linewidth]{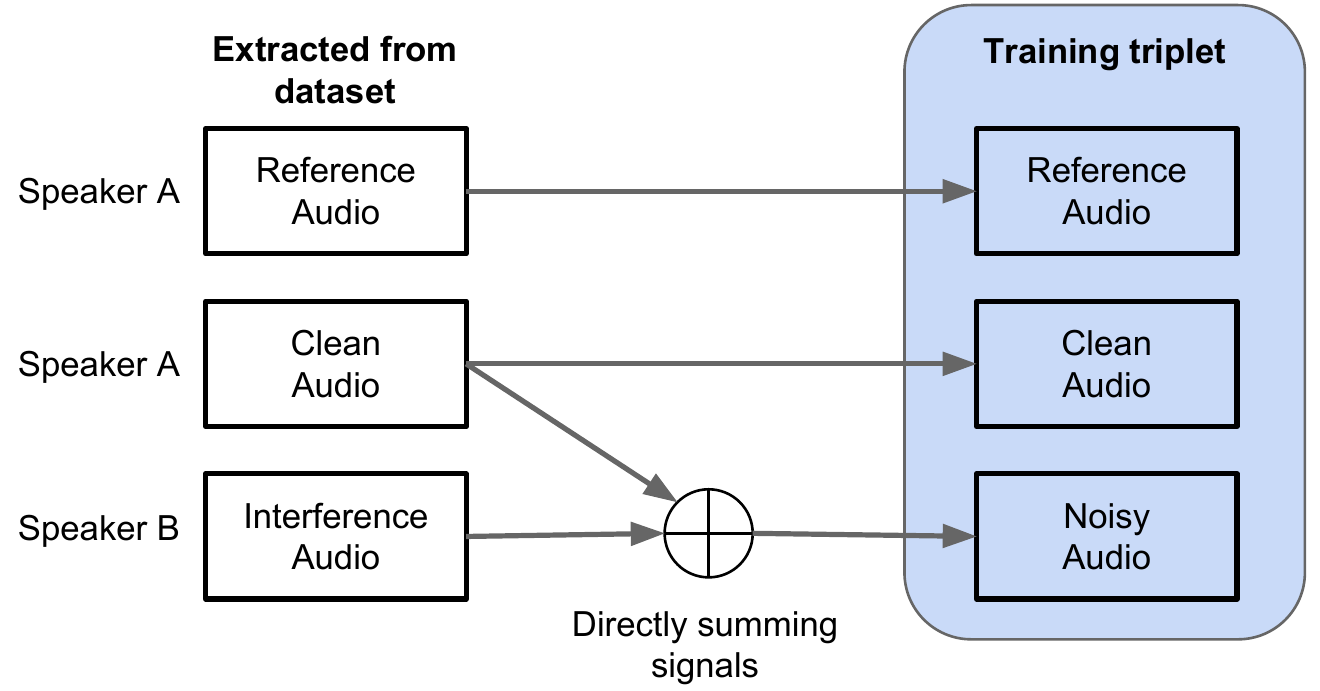}
  \caption{Input data processing workflow.}
  \label{fig:input}
  \vspace{-15pt}
\end{figure}
From the system diagram in Fig.~\ref{fig:system}, we see that one training step involves three inputs: (1) the clean audio from the target speaker, which is the ground truth; (2) the noisy audio containing multiple speakers; and (3) a reference audio from the target speaker (different from the clean audio) over which the d-vector will be computed.

This training triplet can be obtained by using three audios from a clean dataset, as shown in Fig.~\ref{fig:input}. The reference audio is picked randomly among all the utterances of the target speaker, and is different from the clean audio. The noisy audio is generated by mixing the clean audio and an interfering audio randomly selected from a different speaker. More specifically, it is obtained by directly summing the clean audio and the interfering audio, then trimming the result to the length of the clean audio.

We have also tried to multiply the interfering audio by a random weight following a uniform distribution either within $[0,1]$ or within $[0,2]$. However, this did not affect the performance of the VoiceFilter system in our experiments.

\vspace{-5pt}
\subsection{Evaluation}
To evaluate the performance of different VoiceFilter models, we use two metrics: the speech recognition Word Error Rate~(WER) and the Source to Distortion Ratio~(SDR).

\subsubsection{Word error rate}
As mentioned in Sec. \ref{sec:intro}, the main goal of our system is to improve speech recognition. Specifically, we want to reduce the WER in multi-speaker scenarios, while preserving the same WER in single-speaker scenarios.
The speech recognizer we use for WER evaluation is a version of the conventional phone models discussed in \cite{soltau2016neural}, which is trained on a YouTube dataset.

For each VoiceFilter model, we care about four WER numbers:
\begin{itemize} 
	\item \underline{Clean WER}: Without VoiceFilter, the WER on the clean audio.
	\item \underline{Noisy WER}: Without VoiceFilter, the WER on the noisy (clean + interence) audio.
	\item \underline{Clean-enhanced WER}: the WER on the clean audio processed by the VoiceFilter system.
	\item \underline{Noisy-enhanced WER}: the WER on the noisy audio processed by the VoiceFilter system.
\end{itemize}

A good VoiceFilter model should have these two properties:
\begin{enumerate} 
	\item Noisy-enhanced WER is significantly lower than Noisy WER, meaning that the VoiceFilter is improving speech recognition in multi-speaker scenarios.
	\item Clean-enhanced WER is very close to  Clean WER, meaning that the VoiceFilter has minimal negative impact on single-speaker scenarios.
\end{enumerate}

\subsubsection{Source to distortion ratio}
The SDR is a very common metric to evaluate source separation systems~\cite{vincent2006performance}, which requires to know both the clean signal and the enhanced signal. It is an energy ratio, expressed in dB, between the energy of the target signal contained in the enhanced signal and the energy of the errors (coming from the interfering speakers and artifacts). Thus, the higher it is, the better.

\vspace{-5pt}
\section{Results}
\label{sec:results}
\subsection{Word error rate}

\begin{table}
\centering
  \caption{Speech recognition WER on LibriSpeech. VoiceFilter is trained on LibriSpeech.}
  \label{tab:wer_librispeech}
  \vspace{-2pt}
  \begin{tabular}{| c | c | c |}
    \hline
    \multirow{2}{*}{\bf VoiceFilter Model} & \bf Clean & \bf Noisy \\ 
    & \bf WER (\%) & \bf WER (\%) \\ \hline \hline
    No VoiceFilter & 10.9 & 55.9 \\ \hline
    VoiceFilter: no LSTM & 12.2 & 35.3 \\ 
    VoiceFilter: LSTM & 12.2 & 28.2 \\ 
    VoiceFilter: bi-LSTM & \textbf{11.1} & \textbf{23.4} \\  \hline
  \end{tabular}
  \vspace{-5pt}
\end{table}

\begin{table}
\centering
  \caption{Speech recognition WER on VCTK. LSTM layer is uni-directional. Model architecture is shown in Table~\ref{tab:architecture}.}
  \label{tab:wer_vctk}
   \vspace{-2pt}
  \begin{tabular}{| c | c | c |}
    \hline
    \multirow{2}{*}{\bf VoiceFilter Model} & \bf Clean & \bf Noisy \\ 
    & \bf WER (\%) & \bf WER (\%) \\ \hline \hline
    No VoiceFilter & 6.1 & 60.6 \\ \hline
    Trained on VCTK & 21.1 & 37.0 \\
    Trained on LibriSpeech & 5.9 & 34.3 \\ \hline
  \end{tabular}
  \vspace{-15pt}
\end{table}

In Table \ref{tab:wer_librispeech}, we present the results of VoiceFilter models trained and evaluated on the LibriSpeech dataset. The architecture of the VoiceFilter system is shown in Table~\ref{tab:architecture}, with a few different variations of the LSTM layer: (1) no LSTM layer, \textit{i.e.}, only convolutional layers directly followed by fully connected layers; (2) a uni-directional LSTM layer; (3) a bi-directional LSTM layer.
In general, after applying VoiceFilter, the WER on the noisy data is significantly lower than before, while the WER on the clean dataset remains close to before.
There is a significant gap between the first and second model, meaning that processing the data sequentially with an LSTM is an important component of the system. Morever, using a bi-directional LSTM layer we achieve the best WER on the noisy data. With this model, applying the VoiceFilter system on the noisy data reduces the speech recognition WER by a relative 58.1\%. In the clean scenario, the performance degradation caused by the VoiceFilter system is very small: the WER is 11.1\% instead of 10.9\%.

In Table \ref{tab:wer_vctk}, we present the WER results of VoiceFilter models evaluated on the VCTK dataset.
With a VoiceFilter model trained also on VCTK, the WER on the noisy data after applying VoiceFilter is significantly lower than before, reduced relatively by 38.9\%. However, the WER on the clean data after applying VoiceFilter is significantly higher. This is mostly because the VCTK training set is too small, containing only 99 speakers.
If we use a VoiceFilter model trained on LibriSpeech instead, the WER on the noisy dataset further decreases, while the WER on the clean data reduces to 5.9\%, which is even smaller than before applying VoiceFilter. This means: (1) The VoiceFilter model is able to generalize from one dataset to another; (2) We are improving the acoustic quality of the original clean audios, even if we did not explicitly train it this way.
 
Note that the LibriSpeech training set contains about 20 times more speakers than VCTK (2338 speakers instead of 99 speakers), which is the major difference between the two models shown in Table \ref{tab:wer_vctk}. Thus, the results also imply that we could further improve our VoiceFilter model by training it with even more speakers.

\vspace{-5pt}
\subsection{Source to distortion ratio}

\begin{table}[t]
\centering
  \caption{Source to distortion ratio on LibriSpeech. Unit is dB. PermInv stands for permutation invariant loss~\cite{yu2017permutation}. Mean SDR for ``No VoiceFilter'' is high since some clean signals are mixed with silent parts of interference signals.}
  \label{tab:sdr_librispeech}
   \vspace{-2pt}
  \begin{tabular}{| c | c | c |}
    \hline
    \bf VoiceFilter Model & \bf Mean SDR & \bf Median SDR \\ \hline \hline
    No VoiceFilter & 10.1 & 2.5 \\ \hline
    VoiceFilter: no LSTM & 11.9 & 9.7 \\ 
    VoiceFilter: LSTM & 15.6 & 11.3 \\ 
    VoiceFilter: bi-LSTM & \textbf{17.9} & \textbf{12.6} \\  \hline
    PermInv: bi-LSTM & 17.2 & 11.9 \\ \hline
  \end{tabular}
  \vspace{-15pt}
\end{table}

We present the SDR numbers in Table \ref{tab:sdr_librispeech}. The results follow the same trend as the WER in Table~\ref{tab:wer_librispeech}. The bi-directional LSTM approach in the VoiceFilter achieves the highest SDR.

We also compare the VoiceFilter results to a speech separation model that uses the permutation invariant loss~\cite{yu2017permutation}. This model has the same architecture as the VoiceFilter system (with a bi-directional LSTM), presented in Table~\ref{tab:architecture}, but is not fed with  speaker embeddings. Instead, it separates the noisy input into two components, corresponding to the clean and the interfering audio, and chooses the output that is the closest to the ground truth, \textit{i.e.}, with the lowest SDR.  This system can be seen as an ``oracle'' system as it knows both the number of sources contained in the noisy signal as well as the ground truth clean signal. As explained in Section~\ref{sec:intro}, using such a system in practice would require to: 1) estimate how many speakers are in the noisy input, and 2) choose which output to select, \textit{e.g.} by running a speaker verification system on each output (which might not be efficient if there are a lot of interfering speakers).

We observe that the VoiceFilter system outperforms the permutation invariant loss based system. This shows that not only our system solves the two aforementioned issues, but using a speaker embedding also improves the capability of the system to extract the source of interest (with a higher SDR).

\vspace{-5pt}
\subsection{Discussions}
\label{sec:discussions}



In Table \ref{tab:wer_librispeech}, we tried a few variants of the VoiceFilter model on LibriSpeech, and the best WER performance was achieved with a bi-directional LSTM. However, it is likely that a similar performance could be achieved by adding more layers or nodes to uni-directional LSTM. Future work includes exploring more variants and fine-tuning the hyper-parameters to achieve better performance with lower computational cost, but that is beyond the focus of this paper.




\vspace{-5pt}
\section{Conclusions and future work}
\label{sec:conclusions}

In this paper, we have demonstrated the effectiveness of using a discriminatively-trained speaker encoder to condition the speech separation task. Such a system is more applicable to real scenarios because it does not require prior knowledge about the number of speakers and avoids the permutation problem. We have shown that a VoiceFilter model trained on the LibriSpeech dataset reduces the speech recognition WER from 55.9\% to 23.4\% in two-speaker scenarios, while the WER stays approximately the same on single-speaker scenarios.

This system could be improved by taking a few steps: (1) training on larger and more challenging datasets such as VoxCeleb 1 and 2~\cite{chung2018voxceleb2}; (2) adding more interfering speakers; and (3) computing the d-vectors over several utterances instead of only one to obtain more robust speaker embeddings. Another interesting direction would be to train the VoiceFilter system to perform joint voice separation and speech enhancement, \textit{i.e.}, to remove both the interfering speakers and the ambient noise. To do so, we could add different noises when mixing the clean audio with interfering utterances. This approach will be part of future investigations.
Finally, the VoiceFilter system could also be trained jointly with the speech recognition system to further increase the WER improvement.

\vspace{-5pt}
\section{Acknowledgements}
\label{sec:ack}

The authors would like to thank Seungwon Park for open sourcing a third-party implementation of this system.\footnote{\url{https://github.com/mindslab-ai/voicefilter}}
We would like to thank Yiteng (Arden) Huang, Jason Pelecanos, and Fadi \mbox{Biadsy} for the helpful discussions.

\bibliographystyle{IEEEtran}
\bibliography{refs}

\end{document}